\documentclass[letterpaper]{article} % DO NOT CHANGE THIS
\usepackage{aaai24}  % DO NOT CHANGE THIS
\usepackage{times}  % DO NOT CHANGE THIS
\usepackage{helvet}  % DO NOT CHANGE THIS
\usepackage{courier}  % DO NOT CHANGE THIS
\usepackage[hyphens]{url}  % DO NOT CHANGE THIS
\usepackage{graphicx} % DO NOT CHANGE THIS
\usepackage{amssymb}
\usepackage{natbib}  % DO NOT CHANGE THIS AND DO NOT ADD ANY OPTIONS TO IT
\usepackage{caption} % DO NOT CHANGE THIS AND DO NOT ADD ANY OPTIONS TO IT
\usepackage{algorithm}
\usepackage{algorithmic}
\usepackage{booktabs}

\usepackage{newfloat}
\usepackage{listings}
\DeclareCaptionStyle{ruled}{labelfont=normalfont,labelsep=colon,strut=off} % DO NOT CHANGE THIS
\floatstyle{ruled}
\newfloat{listing}{tb}{lst}{}
\floatname{listing}{Listing}
\title{Evaluating the Effects of AI Directors for Quest Selection}
\author {
    Kristen K. Yu,
    Matthew Guzdial,
    Nathan R. Sturtevant
}
\affiliations {
    Computing Science Department, University of Alberta\\
    Alberta Machine Intelligence Institute\\
    \{kkyu, guzdial, nathanst \}@ualberta.ca
}

\begin{document}

\maketitle

\begin{abstract}
Modern commercial games are designed for mass appeal, not for individual players, but there is a unique opportunity in video games to better fit the individual through adapting game elements.  In this paper, we focus on AI Directors, systems which can dynamically modify a game, that personalize the player experience to match the player's preference. In the past, some AI Director studies have provided inconclusive results, so their effect on player experience is not clear. We take three AI Directors and directly compare them in a human subject study to test their effectiveness on quest selection. Our results show that a non-random AI Director provides a better player experience than a random AI Director.
\end{abstract}

\section{Introduction}

 %Outline 
 % 1. AIDs are deployed in commercial games 
 % 2. We can assume that they are having an impact on the player experience 
 % 3. The impact on that player experience is unknown becuase we hvae inconclusive results in several AID evaluations  
 % 4. We want to know what the impact of AIDs on player experience is 
 % 5. We decided to test two AID with inconclusive results to try to gain conclusive results 
 % 6. We compare these AID to a random AID 
 % 7. Random is commonly used in industry becuase it is easy to implement and straightforward to understand 

AI Directors (AID) \cite{AID_L4D} - also called player-centric game AI \cite{charles_player-centred_2005}, experience management \cite{Thue_Dissertation_2010} or drama managers \cite{yu_data-driven_2013} - have been successfully deployed in commercial games. The most famous example of an AID is in Left 4 Dead \cite{L4D}, where the AID maintains a stressful experience for the player by dynamically changing the number of enemies, health packs, and ammunition available in the level \cite{AID_L4D}. A natural question arises from this example: ``What was the impact of the AID on the player experience?". For the Left 4 Dead AID, we do not know because we do not have comparative data on the player experience with their AID, no AID, or other AIDs. Although we do not conclusively know the impact in Left 4 Dead, we can begin to study the impact of AIDs on player experience in other games.  

We may assume there is some positive impact on player experience because AIDs are present in many commercial games \cite{thompson_directors_2017}. Of the studies on AIDs, some have concluded that the AIDs are effective at manipulating the player experience to achieve their desired results \cite{yannakakis_real-time_2009, nygren_userpreference-2011}, while other studies have provided inconclusive results \cite{thue_player-informed_2007, yu2022director}. We target these inconclusive AIDs for further evaluation because we want to better understand the strengths and weaknesses of these AIDs. We can consider the impact across two facets: the first being quantitative changes to player behavior, and the second being qualitative evaluations of the players' perception of their experience. The goal of this paper is to design an experiment to evaluate AIDs along these two facets.

One crucial decision for our experiment is choosing the game in which to evaluate the AIDs. Based on previous studies and examples, AIDs are generally designed for a specific game \cite{thue_interactive_2007, dias_adapting_2011}. This suggests each game requires a unique AID. To address this problem, we choose to implement an AID for a specific system in a game, rather than the complete game. This allows the application of the AID to generalize to any game that uses this system, and we may assume that the findings of the evaluation to hold to other similar games. Specifically, we are interested in quest systems where the player has an option to choose between several quests at a time. We call this problem quest selection, and this design pattern is present in games such as in the Nook Miles+ system in Animal Crossing: New Horizons \cite{acnh}.

For the experiment we used FarmQuest \cite{yu2022testbed}, a video game test bed for AIDs. We compared PaSSAGE \cite{thue_interactive_2007} and a reinforcement learning based AID \cite{yu2022director} to a random algorithm. We chose PaSSAGE and the reinforcement learning AID because they had previously inconclusive results. We chose to include a random AID because it is commonly used in commercial games due to its ease of implementation \cite{yu2022director}. We evaluated these AIDs in a human subject study. 

Our contributions are two-fold. First, we present an evaluation of previously inconclusive AIDs to further characterize their use. Second, to the best of our knowledge, this is the first attempt at directly comparing two existing academic AIDs on the quest selection problem. Our results demonstrate that there is a quantifiable difference in how players play the game with different AIDs, and there is one measured qualitative difference in how players perceive the differences in playing the game.  From these findings, we conclude that a curated AID, either PaSSAGE or the reinforcement learning based AID, performs better on the quest selection problem than random.

\section{Background}

In the past, there have been successful evaluations of AIDs using human subject studies \cite{Wauck_datadriven_2017, yun2010pads} where the authors concluded that their approach was effective at meeting their target goals. However there have been other studies where the authors found their results inconclusive. In this background, we discuss a few specific AIDs which have inconclusive results.

In the study by \citeauthor{thue_interactive_2007}, they proposed PaSSAGE, an AID that changes the story based on the player's previous actions \cite{thue_interactive_2007}. Ninety students participated in a human subject study, and they evaluated the hypothesis that PaSSAGE would be more fun and provide more agency to the players than a non-adaptive version of the story. They only collected survey data. They found that female participants rated PaSSAGE higher in fun and agency, with confidence levels of 93$\%$ and 86$\%$. They identified these results as inconclusive, as other subgroups did not rate PaSSAGE higher, or had a low statistical confidence interval.  

In the study by \citeauthor{dias_adapting_2011}, they implemented a rule-based AID designed to change narrative and other content based on player personality types \cite{dias_adapting_2011}. Twenty-one males were selected for participation such that the different personality types were roughly equal. Within these groups, they compared an adaptive to a non-adaptive version of the game. They collected telemetry data and survey data for their game, such as difficulty, pace, and immersion. They concluded that their AID showed promise, though all but the immersion metric produced inconclusive results. We did not choose this AID for this experiment because it requires participants to take a Myers-Briggs personality test.  

In our previous study, we implemented a reinforcement learning AID that changed which quests to select for a player based on previously accepted quests \cite{yu2022director}. The reinforcement learning AID was compared to a random AID. 208 players played the game, and telemetry data was collected such as time played in game and number of quests presented, accepted, and completed by the player. Survey data was also collected but could not be linked to the AID that the player experienced, so strong conclusions could not be drawn from this data. We concluded that the AID had inconclusive results, because the acceptance rate of quests was not statistically significant, and players spent more time playing the game with the random AID.

\section{Experimental Setup}
In this section we cover the necessary details to understand our experiment. This includes a formal definition of the quest selection problem, a description of the test bed, and a description of the AID used in the experiment. 

\subsection{Quest Selection Problem}

Quest systems in games take on many forms. One common design pattern is to present the player with a few quests at a time. Skyrim \cite{Skyrim} and Animal Crossing: New Horizons \cite{acnh} are two examples of games that have these kind of systems. 

We formalize the quest selection problem as follows. In the game, a player $p$ repeatedly sees $n$ quests $q$ at the same time, and there is a maximum number of quests $m$ that $p$ can have at the same time. All $q$ come from the set $Q$, and $q$ cannot be repeated. $p$ views $\langle q_1, q_2, ..., q_n \rangle$ at the same time. $p$ chooses which of these quests to accept, up to $m$ quests. An AID $d$ selects which $q$ should be shown to $p$, and selects $\langle q_1, q_2, ..., q_n \rangle$.

\subsection{FarmQuest}
\begin{figure}
    \centering
    \includegraphics[width=\columnwidth]{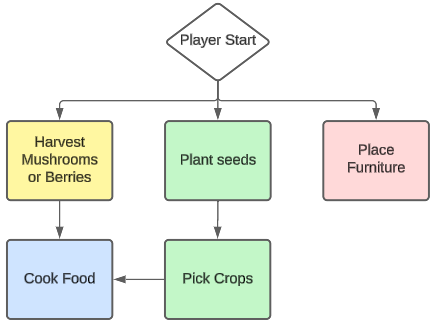}
    \caption{The four main areas of gameplay are placing furniture, planting crops, harvesting, and cooking.}
    \label{fig:pwr_gameloop}
\end{figure}

\begin{figure}
    \centering
    \includegraphics[width=\columnwidth]{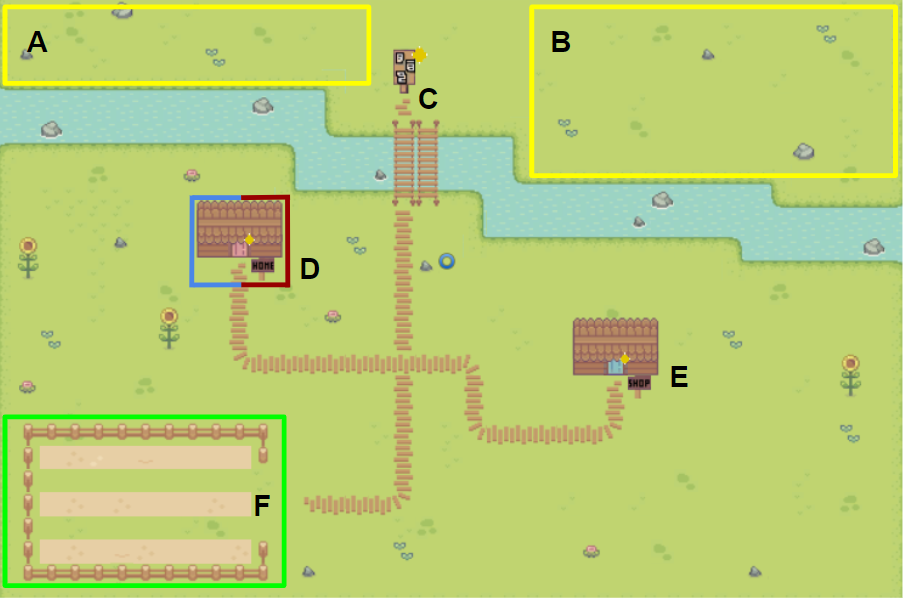}
    \caption{An overview of the main level in FarmQuest}
    \label{fig:pwr_overview}
\end{figure}

FarmQuest is a research test bed video game for AIDs \cite{yu2022testbed}. The game loop consists of planting crops, harvesting mushrooms and berries, placing furniture, and cooking recipes, shown in Figure \ref{fig:pwr_gameloop}. Figure \ref{fig:pwr_overview} shows the main level where players access the different types of gameplay. The region labeled ``A'' is where berries can be harvested, and the region labeled ``B" is where mushrooms can be harvested. The region labeled ``F" is where seeds can be planted, and the region labeled ``D" is where furniture can be placed and recipes can be cooked. There is a shop to buy and sell items in the game, labeled ``E", and a quest board, labeled ``C", where a player can submit or accept quests. The goal is to earn enough coins to pay off their mortgage. The player can earn coins by selling items in the shop or by completing quests. The player starts with 300 coins, and needs to earn 1000 coins to finish the game. 

\begin{figure}
    \centering
    \includegraphics[width=\columnwidth]{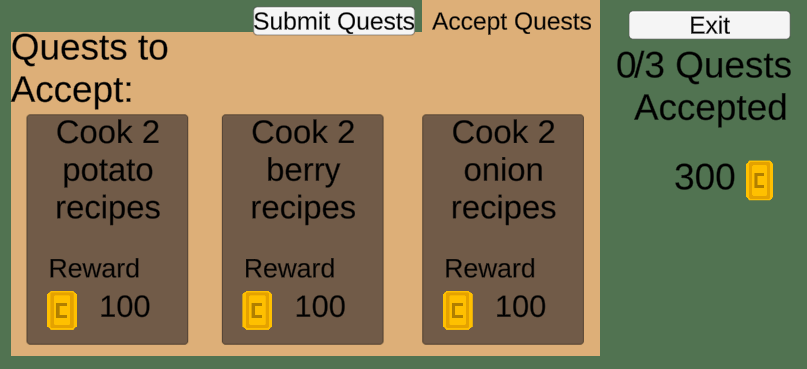}
    \caption{The Questboard that was shown to the players}
    \label{fig:questboard}
\end{figure}

The FarmQuest AID is intended to be embedded in the quest system, where the AID changes which quests are shown to the player. The quest board, shown in Figure \ref{fig:questboard}, allows the player to interact with the quest system. There are two tabs, a submit quest tab and an accept quest tab, and the quest board starts on the submit tab. In the accept quest tab, the current AID selects three quests to present to the player. We will discuss the AIDs in Section 3.3. The player accepts a quest by clicking on it, and can have a maximum number of three quests at a single time. Every time the accept tab is clicked by the player, new quests are chosen by an AID. Once a player completes a quest, the quest can be submitted in the submit tab. Accessing the submit tab of the quest board does not prompt the AID for new quests. 

Each quest is associated with a gameplay type that can be completed using systems in the game. Each quest gives the same reward of 100 coins, to ensure that the reward isn't influencing the decision of the player to accept a quest. There are six place quests, six plant quests, eight cook quests, and six harvest quests as shown in Table \ref{table: example quests}. There are at least six of each quest because that is the minimum number of quests that is needed for the quest system to function. This covers the case where the player accepted three of the same quest type, and the AID selects three more of that quest type to present to the player. There are eight cooking quests in order to cover the breadth of gameplay items that can be cooked.  For the purposes of this paper, each quest is labeled according to the main gameplay type of the quest, where place furniture quests are labeled ``F", plant crop quests are labeled ``P", cook recipe quests are labeled ``C", and harvest berries and mushroom quests are labeled ``H".

\begin{table*}[]
\small
\centering
\begin{tabular}{*4c} 
\hline
Place Quests                                                                                                                                                   & Plant Quests                                                                                                                                              & Cook Quests                                                                                                                                                                                                                                  & Harvest Quests                                                                                                                                                       \\ \hline
\begin{tabular}[c]{@{}l@{}}\textbf{F1} Place 2 furniture \\ \textbf{F2} Place 3 furniture\\ \textbf{F3}  Place 4 furniture\\ \textbf{F4} Place 5 furniture\\\textbf{F5} Place 6 furniture\\\textbf{F6} Place 7 furniture\end{tabular} & \begin{tabular}[c]{@{}c@{}}\textbf{P1}Plant 3 carrots\\\textbf{P2} Plant 3 green onions \\\textbf{P3} Plant 3 lettuce\\ \textbf{P4} Plant 3 onions \\\textbf{P5} Plant 3 potatoes\\\textbf{P6} Plant 3 tomatoes\end{tabular} & \begin{tabular}[c]{@{}c@{}}\textbf{C1}Cook 2 berry recipes\\\textbf{C2} Cook 2 carrot recipes\\\textbf{C3} Cook 2 green onion recipes\\\textbf{C4} Cook 2 lettuce recipes\\\textbf{C5} Cook 2 mushroom recipes \\\textbf{C6} Cook 2 onion recipes\\\textbf{C7} Cook 2 potato recipes\\\textbf{C8} Cook 2 tomato recipes\end{tabular} & \begin{tabular}[c]{@{}c@{}}\textbf{H1} Harvest 3 berries\\\textbf{H2} Harvest 4 berries\\\textbf{H3} Harvest 5 berries\\\textbf{H4} Harvest 3 mushrooms\\\textbf{H5} Harvest 4 mushrooms \\\textbf{H6} Harvest 5 mushrooms\end{tabular} \\ \hline
\end{tabular}
\caption{All of the quests in FarmQuest}
\label{table: example quests}
\end{table*}

\subsection{AI Directors}

There are three AIDs used in this experiment. The first AID is a random algorithm. A common way to implement this quest system is to randomly choose a quest from a pool of existing quests \cite{yu2022director}. However, this comes at the cost of the player experience. Players may be presented with something they do not want to do, which may reduce their motivation to complete the quest. Players may also be asked to undertake a task they have recently accomplished independent of a quest, which may introduce fatigue by completing a quest with those actions. The random algorithm selects a quest from the pool of total quests with a uniform random probability. In our context, it is slightly more likely to select cook quests because there are more cook quests. 

The second AID is PaSSAGE \cite{thue_interactive_2007}. PaSSAGE was originally designed to modify quests in a branching narrative to better suit the player preference. We have adapted it to fit quest selection. PaSSAGE models the player's preference for gameplay based on the actions they have previously taken, and assumes that if a player is taking the action they are enjoying that action. Our implementation of PaSSAGE tracks the number of times a player completes one of the four main gameplay types to use as a player model. PaSSAGE uses rollouts to estimate player return, and determines which of the quests are most appropriate to show to the player. The rollouts stop when there is a decision point - in our case, each rollout is a singular step because there is a singular decision point. PaSSAGE then predicts which type of quest is the most suitable for the player. From there, a specific quest is chosen with uniform random probability from the set of possible quests for the predicted type. 

The third AID is the reinforcement learning AID, which uses a combinatorial multi-armed bandit (CMAB) \cite{yu2022director}. From this point forward, we refer to this AID as the CMAB AID. This AID was originally deployed on a similar quest system to the one in FarmQuest, so the only modifications were the ones necessary to transfer it to the FarmQuest domain. 
Instead of a traditional bandit algorithm where each quest would be an arm, the CMAB AID uses a set of quests as an arm. As shown in Figure \ref{fig:questboard}, there are three quests presented to the player at the same time, and this set of presented quests is the arm for the CMAB AID. CMAB generates the set of all possible quest arms based on the quest type, rather than the set of all quests. For example, a place quest, a cook quest, and a plant quest is a valid arm, but ``place 2 furniture", ``cook 2 onion recipes" and ``plant 3 carrots" is not a valid quest arm. The CMAB AID rewards super arms. A super arm is the set of all arms that have at least one quest type in common, and when one arm is rewarded, all of the arms in the super arm are rewarded. The reward is assigned by the player accepting or not accepting quests, where an accepted quest gives a reward of 1 and an unaccepted quest gives a reward of 0. Then, a particular arm in a super arm is selected using UCB, which gives a set of three quest types. The individual quest is then selected using a uniform random probability from the pool of quests of a given type. 

Given these three AIDs, we hypothesized that the CMAB AID will perform the best on quest selection. The random algorithm has no curation, which might cause the previously outlined friction in the player experience. PaSSAGE assumes that a player's previous actions are an indicator for which quests they will prefer, but players may take actions for various reasons. CMAB assumes that accepting a quest is an indicator of which quests they will prefer. We believe that accepting a quest is a stronger indicator of what a player will prefer than the player's previous actions.

\section{Methodology}

Our AID experiment was an AB test. Though we have three AIDs, we chose to not have participants experience all three directors. Instead, each participant was randomly assigned two out of three of the AIDs in order to shorten the study and reduce confusion involved in comparing three AIDs.  

\begin{figure*}[ht]
    \centering
    \includegraphics[width=\textwidth]{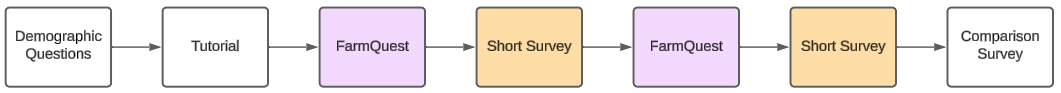}
    \caption{The flow of the participant through the study}
    \label{fig:studyflow}
\end{figure*}

We show the flow of the experimental setup in Figure \ref{fig:studyflow}. First, we asked each player to sign a consent form. Then, the player filled out a demographic survey and completed a short tutorial to learn how to play the game. Then, they played FarmQuest with their first AID until they paid off their mortgage, and answered a short survey about their experience. Then, they played FarmQuest again but with a different AID, and answered the same short survey again. Finally, each player answered a survey comparing the two experiences. The entire experiment took one hour or less to complete. 

The demographic questions were as follows: 

\begin{itemize}
    \item \textbf{DQ1} What is your current age? 
    \item \textbf{DQ2} What is your gender? 
    \item \textbf{DQ3} Do you consider yourself cisgender or transgender?
    \item \textbf{DQ4} Do you consider yourself a gamer? 
    \item \textbf{DQ5} How often do you play games in a week? 
    \item \textbf{DQ6} What kind of genres of games do you play the most? Select all that apply. 
\end{itemize}

Each demographic question is labeled ``DQ" for ease of reference. These questions were all asked because they are considered factors that could affect the player's perception of video games. Age \cite{whitbourne2013reasons}, gender \cite{desai_gender_2017, phan_examining_2012}, and familiarity with video games \cite{Manero_gaminghabits_2017} all could have an impact.

The short survey questions were targeted to the participant's most recent playthrough. We based this survey off of our previous study questions \cite{yu2022director}. We asked players to think about their most recent experience, and to answer the questions accordingly. The short survey questions were all Likert questions on a scale from one to five, where one is strongly disagree and five is strongly agree. The short survey questions were as follows: 

\begin{itemize}
    \item \textbf{SQ1} I felt like I was able to accept quests that I wanted to do 
    \item \textbf{SQ2} I felt like I was \textbf{not} able to accept quests that I wanted to do 
    \item \textbf{SQ3} I felt like I was able to complete quests that I wanted to do 
    \item \textbf{SQ4} I felt like I was \textbf{not} able to complete quests that I wanted to do 
    \item \textbf{SQ5} I felt like there was a variety of quests for me to complete 
    \item \textbf{SQ6} I feel like I enjoyed playing the game
    \item \textbf{SQ7} I feel like I did \textbf{not} enjoy playing the game 
    \item \textbf{SQ8} I feel like I would recommend playing this game to a friend 
    \item \textbf{SQ9}I feel like I would \textbf{not} recommend playing this game to a friend
\end{itemize}

Each short survey question is labeled with ``SQ" for ease of reference. We asked SQ1 and SQ2 to learn about accepting quests, SQ3 and SQ4 to learn about completing quests, and SQ5 to learn about variety. We asked SQ6, SQ7, SQ8 and SQ9 to learn about enjoyment, as it can have an effect on the perception of the AID \cite{yu2022director}. 

The comparison survey questions compared the experience between the two game sessions. We developed the comparison survey questions out of the short survey questions. We asked players to think back on their first experience and compare it to their second experience. 
We first asked a series of Likert questions, using the same scale as the short survey questions. The Likert Questions were as follows: 

\begin{itemize}
    \item \textbf{CQ1} I preferred my first experience playing the game 
    \item \textbf{CQ2} I preferred my second experience playing the game 
    \item \textbf{CQ3} I felt like I accepted more quests that I wanted to do in the first experience 
    \item \textbf{CQ4} I felt like I accepted more quests that I wanted to do in the second experience 
    \item \textbf{CQ5} I felt like I completed more quest that I wanted to do in the first experience 
    \item \textbf{CQ6} I felt like I completed more quests that I wanted to do in the second experience 
    \item \textbf{CQ7} I felt like I had more fun in the first experience 
    \item \textbf{CQ8} I felt like I had more fun in the second experience
\end{itemize}

\noindent
We then asked the players two short answer questions: 

\begin{itemize}
    \item \textbf{CQ9} What was your favorite activity to do in the game? Did you feel like you got to experience a lot of that activity? Why or why not?
    \item \textbf{CQ10} Did you feel that there was a difference between your first and second experiences? Why or why not? 
\end{itemize}

Each comparison survey question is labeled with ``CQ" for ease of reference. We asked pairs of questions to focus on different aspects of player perception. We asked CQ1 and CQ2 for preference, CQ3 and CQ4 for quest acceptance, CQ5 and CQ6 for quest completion, and CQ7 and CQ8 for enjoyment. We asked CQ9 and CQ10 so players could describe their experience in their own words. We asked CQ9 to determine if players would identify as a particular player type. We asked CQ10 to determine if players could tell a difference between AIDs. 

This study was approved by the Research Ethics Board at the University of Alberta (UofA), ethics ID number (Pro00129326). To advertise for the study, we posted digital calls for participation in classroom forums and Discord at the UofA, and AI and video game studio Slack spaces. The only requirement for participation was that a person be over the age of 18. There was no financial incentive.

\section{Results}
In this section, we discuss the demographic, qualitative, and quantitative results of the human subject study. 

\subsection{Demographic Results}
First, we present the demographic results in order to gain an understanding of any potential biases from the audience that participated in this study. In total, 108 players started the study. Of these, only thirty-nine players fully completed the study. Most players stopped the study after filling out the demographic questions. Of the thirty-nine complete responses, seven of these included data where the players refreshed the page at some point, which caused the system to assign a different AID. This means that we cannot know which experience they are talking about when they answered the survey questions, so we do not include this data. We are left with thirty-two complete responses that have usable data.

Table \ref{tab:demographics_age_gender} shows the self-reported age and gender of players. Twenty-nine players identified as cisgender, two identified as transgender, and one did not answer.  Table \ref{tab:demographics_genre} shows the genres of games played by the players, where each player could select more than one option. For whether a player considered themselves a gamer or not, twenty-eight reported yes and four reported no. Table \ref{tab:demographics_timeplayinggames} shows the self-reported number of times the participants played games in a week.

\begin{table}[h]
\small
\centering
\begin{tabular}{@{}lc|lc@{}}
\toprule
Reported Age & Players & Reported Gender   & Players \\ \midrule
18-25        & 7       & Man               & 23      \\
26-35        & 16      & Woman             & 6       \\
36-45        & 8       & Non-Binary        & 2       \\
45 or older  & 1       & Prefer not to say & 1       \\ \bottomrule
\end{tabular}
\caption{The reported age and gender of players}
\label{tab:demographics_age_gender}
\end{table}

\begin{table}[h]
\small
\centering
\begin{tabular}{@{}lc|lc@{}}
\toprule
Genre     & Players & Genre      & Players \\ \midrule
Puzzle    & 17      & Platformer & 12      \\
Arcade    & 7       & Sports     & 3       \\
RPG       & 25      & Racing     & 6       \\
FPS       & 20      & Simulation & 16      \\
Action    & 20      & Fighting   & 2       \\
Adventure & 26      & Other      & 5      \\ \bottomrule
\end{tabular}
\caption{The genres of games that the players played}
\label{tab:demographics_genre}
\end{table}

\begin{table}[]
\centering
\small
\begin{tabular}{@{}ll@{}}
\toprule
Reported time spent playing games & Players \\ \midrule
Less than once a week             & 1       \\
Once a week                       & 5       \\
Two or three times a week         & 3       \\
Four or five times a week         & 10      \\
Six or seven times a week         & 13      \\ \bottomrule
\end{tabular}
\caption{The reported amount of time spent playing games in a week}
\label{tab:demographics_timeplayinggames}
\end{table}

Table \ref{tab:numplayers-aid-order} shows the number of players and the specific ordering of AIDs each player experienced. In total, twenty-two players experienced PaSSAGE, twenty experienced CMAB, and twenty-two experienced random. 

\begin{table}[]
\centering
\small
\begin{tabular}{@{}lc@{}}
\toprule
AID Ordering   & Number of Players \\ \midrule
PaSSAGE then Random & 5                 \\
PaSSAGE then CMAB   & 3                 \\
CMAB then PaSSAGE   & 7                 \\
CMAB then Random    & 3                 \\
Random then PaSSAGE & 7                 \\ 
Random then CMAB    & 7                 \\ \bottomrule
\end{tabular}
\caption{The number of players for each ordering of AIDs}
\label{tab:numplayers-aid-order}
\end{table}

\subsection{Quantitative Results}
We present the quantitative results from the study. First, we present the number of quests presented, accepted, and completed by each player, shown in Table \ref{tab:quest_data}. We wanted to see if the order of sessions had an effect on any of this data, so we ran a Kruskal-Wallis statistical test. None of these tests yielded statistically significant results, so we assume that ordering did not have an effect and combined the data. We wanted to see if there is a difference in any of this data based on AID, so we ran Mann-Whitney U tests. Only the number of presented quests had statistically significant results, shown in bold in Table \ref{tab:quest_data}. The CMAB AID had statically significantly fewer presented quests than both PaSSAGE ($p = 0.032$) and random ($p = 0.047$) AIDs.

Second, we present the average playtime. To calculate the time spent in game, we cannot use the timestamp data included in our telemetry due to a problem with the logging. Instead, we use the number of days spent in game as an approximation. In game, each day is thirty seconds, each twilight is ten seconds and each night is twenty seconds, so a complete in-game day is one minute. Thus, we can measure the time spent in game to floor of the nearest minute.

Table \ref{tab:day_data_total} shows the average, stdev, and standard error of the mean (SEM) on the number of in game days played by players. The SEM is a measure of how far the sample mean is from the population mean. To see if ordering of session had an effect on the duration that players played the game, we ran a Kruskal-Wallis statistical test. Only the random AID had a statistically significant result ($p = 0.011$) with a 95 $\%$ confidence interval, shown in bold in Table \ref{tab:day_data_total}. This means that ordering of session does not have an effect in the cases of the PaSSAGE and CMAB AIDs, but does have an effect with a random AID. For the random AID, we wanted to verify if the playtime for session 2 is statistically lower. We ran a Mann-Whitney U test with the alternate hypothesis that session 2 is greater than session 1. This showed a statistically significant result with a 95 $\%$ confidence interval ($p = 0.006$), so session 1 is statistically longer than session 2. This means that players who were assigned the random AID in session 1 played statistically longer than the players who were assigned the random AID in session 2.

\begin{table}[]
\centering
\small
\begin{tabular}{@{}lccc@{}}
\toprule
AID                                                  & \begin{tabular}[c]{@{}c@{}}Average In \\ Game Days\end{tabular} & Stdev & SEM\\ \midrule
\begin{tabular}[c]{@{}l@{}}Session 1 Passage\end{tabular} & 8.75                                                            & 4.20                                                          & 1.48                                                       \\
\begin{tabular}[c]{@{}l@{}}Session 2 Passage\end{tabular}  & 8.35                                                            & 6.36                                                          & 1.70                                                       \\
\begin{tabular}[c]{@{}l@{}}Session 1 CMAB\end{tabular}    & 12.30                                                           & 4.72                                                          & 1.49                                                       \\
\begin{tabular}[c]{@{}l@{}}Session 2 CMAB\end{tabular}    & 8.70                                                            & 4.90                                                          & 1.54                                                       \\
\begin{tabular}[c]{@{}l@{}}Session 1 Random\end{tabular}   & \textbf{11.28}                                                           & 5.61                                                          & 1.49                                                       \\
\begin{tabular}[c]{@{}l@{}}Session 2 Random\end{tabular}   & \textbf{7.25}                                                            & 6.48                                                          & 2.29                                                       \\ \bottomrule
\end{tabular}
\caption{The average number of in games days per AID and session, where bold indicates statistically significant results}
\label{tab:day_data_total}
\end{table}

\begin{table*}[]
\centering
\small
\begin{tabular}{lccc|lccc|lccc} \toprule
Presented & Average & Std Dev & SEM  & Accepted & Average & Std Dev & SEM  & Completed & Average & Std Dev & SEM  \\ \midrule
PaSSAGE   & \textbf{27.27}   & 13.85   & 2.95 & PaSSAGE  & 8.59    & 3.40    & 0.72 & PaSSAGE   & 7.09    & 3.61    & 0.77 \\
CMAB      & \textbf{20.68}   & 8.36    & 1.92 & CMAB     & 7.89    & 1.88    & 0.43 & CMAB      & 6.58    & 2.01    & 0.46 \\
Random    & \textbf{29.71}   & 22.06   & 4.81 & Random   & 8.24    & 2.96    & 0.65 & Random    & 7.05    & 2.65    & 0.58 \\ \bottomrule
\end{tabular}
\caption{The number of presented, accepted and completed quests by AID, where bold indicates statistically significant results}
\label{tab:quest_data}
\end{table*}

\subsection{Qualitative Results}
We present the qualitative results from the study. First, we present the results from questions SQ1-SQ9. We wanted to determine if the ordering of the AIDs had an effect. We ran a Kruskal-Wallis statistical test on the responses, where a statistically significant result would show that ordering does have an effect. Only SQ6, ``I feel like I enjoyed playing the game" had a statistically significant difference for the random AID with a 95$\%$ confidence interval ($p = 0.028$). This means we can treat the SQ6 answers from sessions without a random AID as coming from the same distribution, so we combined the data. Table \ref{tab:short_survey} shows the results from the short survey with SQ6 for random removed. 

The median result for session 1 of the random AID is 4.00, and the median result for session 2 of the AID is 2.50. The average result for session 1 of the random AID was 3.61, and the average result for session 2 of the random AID was 2.12. According to the Likert scale, a higher number means that they agree more with the statement ``I feel like I enjoyed playing the game". 

To determine if the results for session 2 are from a lower distribution, we ran a Mann Whitney-U test on the results from SQ6 with the alternate hypothesis that session 1 is greater than session 2. We ran this test on all three AIDs. The PaSSAGE ($p = 0.765$) and CMAB ($p = 0.578$) AIDs were not statistically significant, but the random AID was statistically significant, indicating less enjoyment ($p = 0.016$). 

\begin{table*}[]
\centering
\small
\begin{tabular}{@{}lccccccccc@{}}
\toprule
AID & SQ1 & SQ2 & SQ3 & SQ4 & SQ5 & SQ6 & SQ7 & SQ8 & SQ9 \\ %\midrule
PaSSAGE     & 4  & 2  & 4  & 2  & 4  & 4  & 3  & 3  & 3  \\
CMAB        & 4  & 2  & 4  & 2  & 4  & 3  & 3  & 2  & 3  \\
Random      & 4  & 2  & 4  & 2  & 3  & -  & 2  & 3  & 3  \\ \bottomrule
\end{tabular}
\caption{Comparison of the median value for short survey questions, where 1 is strongly disagree and 5 is strongly agree}
\label{tab:short_survey}
\end{table*}

The results from the comparison questions are shown in the appendix because there was no statistical significant with a 95$\%$ confidence interval.

\section{Discussion}
In this section, we discuss the two facets of the impact of AIDs: quantitative changes in behavior and player's perception of impact. 

\subsection{Quantitative Changes in Behavior}
We hypothesized that the CMAB AID would be the most suitable for this particular game and the quest selection problem. However, our results do not show that the CMAB AID is the clear winner, and instead showed that either the PaSSAGE or CMAB AID is preferable to players. We looked at the number of quests presented, accepted, and completed by the player, the total play time, and the short survey data to support this claim. 

The CMAB AID had statistically significantly fewer presented quests than both the Random and PaSSAGE AIDs, while still having a similar number of accepted and completed quests. This shows that players were able to accept and complete similar numbers of quests with all three AIDs, while the CMAB AID needed to present the fewest number of quests to do so. This shows that the CMAB AID was effectively able to reduce time spent searching for quests.

The total playtime shows differences in how long players played, but is not statistically significant, as shown in Table \ref{tab:day_data_total}. The playtime for CMAB trends higher than random in both sessions, which suggests that the CMAB AID is able to perform well in both session 1 and session 2 for the player. The playtime for PaSSAGE is lower than the random AID during session 1. We compared PaSSAGE session 2 to random session 2, and PaSSAGE trended higher than the random AID. However, this result was also not statistically significant. This data shows there may be a preference for either PaSSAGE or the CMAB AIDs. 

In the short survey data, SQ6 had a statistically significant result. Players indicated that they did not enjoy playing random during the second session, shown in Section 5.3. In this experiment, they experienced either the PaSSAGE or CMAB AIDs first, where AIDs curated the quests to the player. We anticipate the difference between a curated and random experience was more apparent when faced with a curated experience first. This data showed a preference in players for either non-random AIDs in terms of enjoyment.

\subsection{Player's Perception of AIDs}
Thus far we have discussed the evidence for how AIDs quantitatively affect the player experience, but the question remains: did players perceive a difference in experience? We conclude that players  under different AIDs will play the game differently, but do not perceive a difference. We discuss the short survey data, the comparison survey data, and some free responses to support this conclusion.

For the short survey data, most of the responses showed no difference between AIDs. Additionally, the comparison questions did not show any statistically significant results, as shown in the appendix. Players did not notice a difference between their experiences for the majority of the qualitative questions, and their responses reflect that. When directly asked in question CQ10, there were some players who responded ``No, I didn't notice any difference", or ``I did not feel like there was a huge difference between the two experiences." However, we had measurable quantitative differences. We conclude that players play differently with curated vs non-curated AIDs, but fail to notice a difference in their play style that is caused by the AID. 

\section{Threats to Validity}

One possible threat to validity to our findings is the effect of salience on players selecting quests. Salience is the idea that some objects, when presented together, are more noticeable by a person based on their ordering. As shown in Figure \ref{fig:questboard}, three quests are presented to the player in a row. There could be a salience effect where native English speakers notice the left-most quest more because English speakers read left to right. To determine if this is the case, we analyzed the placement data of each quest according to quest type, shown in the appendix. We ran a Chi-Squares contingency test on the number of presented quests in each position compared to the number of accepted quests in each position, and we found that there was no statistically significant results with a confidence interval of 95$\%$. We conclude that salience did not have an effect on players selecting quests.

Another possible threat to validity is the effect of difficulty on selecting quests. It could be that players are selecting quests based on what is easiest. To address this, we analyzed the difficulty within each quest type. For place and harvest quests, we assume that a smaller number of items placed or harvested means an easier quest. For cook quests, we assume that C1 and C5 are easier because the ingredients do not require planting, and assume that all other cook quests are of equal difficulty because they take the same number of actions to complete. For plant quests, we assume that all quests have an equal difficulty because they all require the same number of actions to complete. We ran a Chi-Squares contingency test on the number of presented and accepted quests for each individual quest, to see if there is a bias towards quests that are easier. There is only one category that had statistical significance, which was the cooking type quests for the random AID ($p value = 0.008$). We believe this is due to the higher number of acceptances of the quest C7. This quest is assumed to be similar difficulty to the other cook quests except for C1 and C5. Because C7 is the quest with a higher number of acceptances, and not C1 or C5, this does not indicate that there are a higher number of people choosing an easier quest. The other statistical tests do not show any significance. Thus, we conclude that the difficulty does not have an effect on players selecting quests.

\section{Conclusion}
In this paper we evaluated previously inconclusive AIDs on quest selection. We directly compared tested these AIDs in a human subject study. We collected both quantitative and qualitative data, which showed that a curated AID leads to a longer time played and less quests presented than a random AID, but players fail to perceive a difference. 

In the future, we hope to gain a larger sample for the comparison between the PaSSAGE and CMAB AIDs, as the small sample size limited some of the analysis on our data. This could help disambiguate which AID is better for this problem, if either. Additionally, we hope to test other AIDs on the quest selection problem.  This will help paint a clearer picture of the individual strengths and weaknesses of existing AIDs, so we can better understand the gaps that need to be addressed in future research. 

\section*{Acknowledgements}
This work was funded by the Canada CIFAR AI Chairs Program, Alberta Machine Intelligence Institute, and the Natural Sciences and Engineering Research Council of Canada (NSERC).

\bibliography{AID_Argument,AI_Directors,Games, QuestDefinitions, GeneralQuestTheory}
\newpage
\section*{Appendix}

Tables \ref{tab:salience_presented_accepted} and \ref{tab:salience_passage_accepted} show the presented and accepted quests by position for the PaSSAGE AID, respectively. Tables \ref{tab:salience_RLAID_presented} and \ref{tab:salience_RLAID_Accepted} show the presented and accepted quests by position for the CMAB AID, respectively. Tables \ref{tab:salience_random_presented} and \ref{tab:salience_random_accepted} show the presented and accepted quests by position for the random AID, respectively.

For the presented quests with the PaSSAGE AID, there is a bias for planting quests to appear in the left position, and harvest quests to appear on the right position. For presented quests with the CMAB AID, there is a similar bias to passage where planting quests most frequently appear on the left position, and harvest quests most frequently appear on the right. 

To ensure that these biases have no affect on which quest the player is selecting, we ran a chi-squares contingency test for each AID and quest type. We compared the amount of presented quests in each position to the amount of quests accepted in each position. None of these tests resulted in a statistically significant result. Thus, we conclude that salience is not a contributing factor in whether a player selects a quest.

% \usepackage{booktabs}
\begin{table}[]
\small
\centering
\begin{tabular}{@{}lccc@{}}
\toprule
PaSSAGE Presented & Left & Middle & Right \\ \midrule
Place             & 30   & 54     & 57    \\
Plant             & 99   & 35     & 11    \\
Cook              & 51   & 59     & 35    \\
Harvest           & 9    & 41     & 86    \\ \bottomrule
\end{tabular}
\caption{The number of quests in each position that were presented to the player using the PaSSAGE AID}
\label{tab:salience_presented_accepted}
\end{table}

% \usepackage{booktabs}
\begin{table}[]
\centering
\small
\begin{tabular}{@{}lccc@{}}
\toprule
PaSSAGE Accepted & Left & Middle & Right \\ \midrule
Place            & 5    & 8      & 9     \\
Plant            & 20   & 9      & 0     \\
Cook             & 13   & 16     & 10    \\
Harvest          & 7    & 22     & 55    \\ \bottomrule
\end{tabular}
\caption{The number of quests in each position that were accepted by the player using the PaSSAGE AID}
\label{tab:salience_passage_accepted}
\end{table}

% \usepackage{booktabs}
\begin{table}[]
\centering
\small
\begin{tabular}{@{}lccc@{}}
\toprule
CMAB Presented & Left & Middle & Right \\ \midrule
Place          & 14   & 28     & 43    \\
Plant          & 59   & 30     & 4     \\
Cook           & 54   & 43     & 25    \\
Harvest        & 10   & 36     & 65    \\ \bottomrule
\end{tabular}
\caption{The number of quests in each position that were presented to the player using the CMAB AID}
\label{tab:salience_RLAID_presented}
\end{table}

% \usepackage{booktabs}
\begin{table}[]
\centering
\small
\begin{tabular}{@{}lccc@{}}
\toprule
CMAB Accepted & Left & Middle & Right \\ \midrule
Place         & 7    & 13     & 16    \\
Plant         & 20   & 11     & 2     \\
Cook          & 14   & 14     & 8     \\
Harvest       & 4    & 18     & 34    \\ \bottomrule
\end{tabular}
\caption{The number of quests in each position that were accepted by the player using the CMAB AID}
\label{tab:salience_RLAID_Accepted}
\end{table}

% \usepackage{booktabs}
\begin{table}[]
\centering
\small
\begin{tabular}{@{}lccc@{}}
\toprule
Random Presented & Left & Middle & Right \\ \midrule
Place            & 57   & 44     & 42    \\
Plant            & 62   & 51     & 63    \\
Cook             & 50   & 76     & 63    \\
Harvest          & 51   & 49     & 50    \\ \bottomrule
\end{tabular}
\caption{The number of quests in each position that were presented to the player using the random AID}
\label{tab:salience_random_presented}
\end{table}

% \usepackage{booktabs}
\begin{table}[]
\centering
\begin{tabular}{@{}lccc@{}}
\toprule
Random Accepted & Left & Middle & Right \\ \midrule
Place           & 11   & 17     & 14    \\
Plant           & 8    & 12     & 12    \\
Cook            & 12   & 11     & 13    \\
Harvest         & 23   & 29     & 23    \\ \bottomrule
\end{tabular}
\caption{The number of quests in each position that were accepted by the player using the random AID}
\label{tab:salience_random_accepted}
\end{table}

We wanted verify that the random algorithm trends towards equal positions in the long term. Thus, we ran an experiment where we tracked the position quests in the random algorithm 10,000 times, and recorded the position in appendix table \ref{tab:unity-experiment-random}. There are twenty-six total quests in game. For the place, plant, and harvest type quests there are six individual quests, which represents 23$\%$ of the total amount of quests. The place, plant, and harvest type quests all have aproximately 2300, or 23$\%$ of 10,000 quests in each position. For the cook type quest, there are eight individual quests, which represents 30$\%$ of the total amount of quests. The cook type quests have approximately 3000, or 30$\%$ of 10,000 quests in each position. This shows there is not a problem with the code, and the higher numbers seen during the AID experiment would no longer be present if more data was collected.

\begin{table*}[]
\centering
\small
\begin{tabular}{@{}lcccccccc@{}}
\toprule
                  & \textbf{C1} & \textbf{C2} & \textbf{C3} & \textbf{C4} & \textbf{C5} & \textbf{C6} & \textbf{C7} & \textbf{C8} \\ \midrule
PaSSAGE Presented & 16          & 16          & 22          & 17          & 19          & 13          & 13          & 16          \\
PaSSAGE Accepted  & 10          & 5           & 3           & 3           & 7           & 2           & 5           & 3           \\
CMAB Presented    & 11          & 13          & 23          & 14          & 16          & 17          & 13          & 15          \\
CMAB Accepted     & 7           & 4           & 3           & 1           & 7           & 4           & 7           & 5           \\
Random Presented  & 11          & 17          & 21          & 15          & 23          & 16          & 16          & 17          \\
Random Accepted   & 7           & 3           & 2           & 4           & 8           & 2           & 10          & 6           \\ \bottomrule
\end{tabular}
\caption{The number of presented and accepted quests for cook type quests}
\label{tab:cook-quest-difficulty}
\end{table*}

\begin{table}[h]
\centering
\small
\begin{tabular}{@{}lccc@{}}
\toprule
Random Experiment & Left & Middle & Right \\ \midrule
Place             & 2340 & 2338   & 2311  \\
Plant             & 2296 & 2319   & 2267  \\
Cook              & 3047 & 3034   & 3039  \\
Harvest           & 2317 & 2309   & 2329  \\ \bottomrule
\end{tabular}
\caption{Position results from of evaluating the random AID 10,000 times}
\label{tab:unity-experiment-random}
\end{table}

To determine if difficulty has any effect, we looked at the presented and accepted quests by quest type. Tables   \ref{tab:cook-quest-difficulty}, \ref{tab:harvest-quest-difficulty},\ref{tab:place-quest-difficulty}, \ref{tab:plant-quest-difficulty} shows the results for all three AIDs of the number of presented and accepted quests.

\begin{table}[]
\small
\centering
\begin{tabular}{@{}lcccccc@{}}
\toprule
& \textbf{H1} & \textbf{H2} & \textbf{H3} & \textbf{H4} & \textbf{H5} & \textbf{H6} \\ \midrule
PaSSAGE Presented & 27                                                           & 25                                                           & 21                                                           & 15                                                             & 32                                                             & 27                                                             \\
PaSSAGE Accepted  & 18                                                           & 17                                                           & 14                                                           & 9                                                              & 13                                                             & 12                                                             \\
CMAB Presented    & 24                                                           & 21                                                           & 19                                                           & 9                                                              & 24                                                             & 23                                                             \\
CMAB Accepted     & 13                                                           & 11                                                           & 10                                                           & 6                                                              & 5                                                              & 10                                                             \\
Random Presented  & 23                                                           & 24                                                           & 12                                                           & 12                                                             & 32                                                             & 18                                                             \\
Random Accepted   & 13                                                           & 16                                                           & 6                                                            & 7                                                              & 14                                                             & 10                                                             \\ \bottomrule
\end{tabular}
\caption{The number of presented and accepted quests for harvest type quests}
\label{tab:harvest-quest-difficulty}
\end{table}

% \usepackage{booktabs}
\begin{table}[]
\small
\centering
\begin{tabular}{@{}lcccccc@{}}
\toprule
                  & \textbf{F1} & \textbf{F2} & \textbf{F3} & \textbf{F4} & \textbf{F5} & \textbf{F6}\\ \midrule
PaSSAGE Presented & 18                                                           & 22                                                           & 17                                                           & 24                                                           & 16                                                           & 26                                                           \\
PaSSAGE Accepted  & 9                                                            & 5                                                            & 2                                                            & 4                                                            & 3                                                            & 0                                                            \\
CMAB Presented    & 15                                                           & 17                                                           & 13                                                           & 9                                                            & 12                                                           & 14                                                           \\
CMAB Accepted     & 9                                                            & 8                                                            & 4                                                            & 2                                                            & 5                                                            & 5                                                            \\
Random Presented  & 21                                                           & 23                                                           & 18                                                           & 27                                                           & 24                                                           & 24                                                           \\
Random Accepted   & 10                                                           & 7                                                            & 4                                                            & 6                                                            & 8                                                            & 5                                                            \\ \bottomrule
\end{tabular}
\caption{The number of presented and accepted quests for place type quests}
\label{tab:place-quest-difficulty}
\end{table}

% \usepackage{booktabs}
\begin{table}[]
\small
\centering
\begin{tabular}{@{}lcccccc@{}}
\toprule
& \textbf{P1} & \textbf{P2} & \textbf{P3} & \textbf{P4}& \textbf{P5} & \textbf{P6} \\ \midrule
PaSSAGE Presented & 23                                                         & 21                                                              & 16                                                         & 22                                                        & 18                                                          & 17                                                          \\
PaSSAGE Accepted  & 4                                                          & 4                                                               & 3                                                          & 9                                                         & 8                                                           & 2                                                           \\
CMAB Presented    & 14                                                         & 11                                                              & 13                                                         & 13                                                        & 18                                                          & 14                                                          \\
CMAB Accepted     & 8                                                          & 2                                                               & 4                                                          & 5                                                         & 9                                                           & 3                                                           \\
Random Presented  & 27                                                         & 20                                                              & 17                                                         & 23                                                        & 22                                                          & 19                                                          \\
Random Accepted   & 6                                                          & 2                                                               & 5                                                          & 8                                                         & 9                                                           & 3                                                           \\ \bottomrule
\end{tabular}
\caption{The number of presented and accepted quests for plant type quests}
\label{tab:plant-quest-difficulty}
\end{table}

In the comparison survey, we asked players to directly compare their experiences to each other. Table \ref{tab:numplayers-aid-order} shows the number of players who had each experience. We performed a kruskal-wallis statistical test on the results of the comparison survey to determine if ordering had an effect. None of the tests showed a statistically significant difference, so the ordering has no effect. This means we can directly compare AIDs, with twelve players for PaSSAGE vs Random, ten for PaSSAGE vs CMAB and nine for Random vs CMAB. 

There was a problem with collecting the information for questions 3 and 4, so there are no results for this data. Because the questions in the comparison survey are directly asking about each experience, we have unobfuscated which director the question was asking about. We aggregated the responses for each AID for which provided a better experience, which AID felt like players were completing more quests, and which AID was more fun. These three comparisons are abbreviated ``experience", ``complete" and ``fun" in all table results. 

Table \ref{tab:passage_vs_random_comparison} shows the median comparison survey results of PaSSAGE and Random AIDs. 
Table\ref{tab:passage_vs_RLAID_comparison} shows the median comparison survey results of PaSSAGE and CMAB AIDs. Table \ref{tab:random_vs_RLAID_comparison} shows the median comparison results of CMAB and Random AIDs. We ran Mann-Whitney U statistical tests on the comparison survey data and none of them yielded statistically significant results.

% \usepackage{booktabs}
\begin{table}[h]
\centering
\small
\begin{tabular}{@{}lccc@{}}
\toprule
PaSSAGE vs Random & Experience & Complete & Fun \\ \midrule
PaSSAGE           & 3          & 2        & 3   \\
Random            & 3          & 4        & 3   \\ \bottomrule
\end{tabular}
\caption{Median Likert comparison responses for PaSSAGE vs random AIDs}
\label{tab:passage_vs_random_comparison}
\end{table}

\begin{table}[h]
\centering
\small
\begin{tabular}{@{}llll@{}}
\toprule
PaSSAGE vs CMAB & Experience & Complete & Fun \\ \midrule
PaSSAGE         & 3          & 3        & 3   \\
CMAB            & 3          & 3.5      & 3   \\ \bottomrule
\end{tabular}
\caption{Median Likert comparison responses for PaSSAGE vs CMAB AIDs}
\label{tab:passage_vs_RLAID_comparison}
\end{table}

\begin{table}[]
\centering
\small
\begin{tabular}{@{}llll@{}}
\toprule
CMAB vs Random & Experience & Complete & Fun \\ \midrule
CMAB           & 3.5        & 3        & 3.5 \\
Random         & 3          & 3        & 3   \\ \bottomrule
\end{tabular}
\caption{Median Likert comparison responses for CMAB vs random AIDs}
\label{tab:random_vs_RLAID_comparison}
\end{table}

\end{document}